\documentclass[epj]{svjour}
\usepackage{latexsym}
\usepackage{graphics}
\begin{document}
\title{Self-Organized Criticality on Quasiperiodic Graphs}

\author{Dieter Joseph
}
\institute{Max-Planck-Insitut f\"ur Physik komplexer Systeme,
N\"othnitzer Str.\ 38, D-01187 Dresden, Germany}
\date{Received: date / Revised version: date}
\abstract{
Self-organized critical models are used to describe the 1/f-spectra of
rather different physical situations like snow avalanches,
noise of electric currents, luminosities of stars or
topologies of landscapes. The prototype of the SOC-models is the sandpile
model of Bak, Tang and Wiesenfeld (Phys.\ Rev.\ Lett.\ {\bf 59}, (1987) 381). 
We implement this model
on non-periodic graphs where it can become either isotropic or anisotropic
and compare its properties with the periodic counterpart on the square
lattice.
} 
\maketitle
\section{Introduction}
\label{intro}
Self-organized criticality (SOC) is observed in various dynamical systems 
that develop
algebraic correlations. The systems drive themselves into critical states
independent of initial states or tuning of parameters. The complex
behaviour is characterized by the absence of characteristic time or
length scales.  It is possible
to define a set of critical scaling exponents that characterize the large scale
behaviour and describe the universal 1/f-power spectra (for an overview
see Ref.\ 2 in \cite{Vespa}).

Sandpile automata \cite{BTW} are among the simplest models that develope SOC 
behaviour in the avalanche propagation.
The model is defined as follows: on each site of a lattice, a variable $z_i$
is defined (number of grains of sand). At each time step, the variable at a
randomly chosen
site is increased $z_i \rightarrow z_i+1$. Above a critical value $z_c$,
the site topples $z_i \rightarrow z_i-z_c$ and the nearest neighbors $j$ of 
site $i$ are increased $z_j \rightarrow z_j+1$. Thereby, $z_c$ grains of sand 
have been transported from site $i$ to the neighbors $j$.
A toppling can induce nearest neighbor topplings and can create avalanches
of arbitrary sizes $s$.
For finite systems, one typically uses
open boundary conditions (periodic boundary conditions can lead to never
ending avalanches), i.e.\ the sand is just dropping from the table
if it reaches the border. This introduces the concept of
finite size scaling (FFS) to compensate cutoff effects. 
Under the assumption that FFS is valid the avalanches
are described by three variables: the number of toppling $s$, the area
$a$ covered by an avalanche and the avalanche duration $T$. The probability
distributions of these variables are algebraic:
\begin{equation}
P(x)=x^{-\tau_x}G(\frac{x}{x_c})
\end{equation}
with $x\in\{s,a,T\}$. $G$ is a cutoff function and the cutoff diverges
$x_c \sim L^{\beta_x}$ with the system size $L \rightarrow \infty$.
The universality class is determined by the set of exponents $\{\tau_x,
\beta_x\}$. Very recently, the validity of FFS for sandpile models
was questioned \cite{Drossel} but this discussion is far beyond the
scope of this article. We refer to \cite{Drossel} and references therein
and take FFS as a working hypothesis hereafter.

The sandpile automata can be isotropic \cite{BTW}, i.e the sand is
distributed equally between nearest neighbors, or anisotropic
\cite{Manna}. The question whether both belong to the same universality
class is still under discussion. Whereas real-space normalization group
\cite{Pietro}, field theoretical \cite{Dickman} and late numerical approaches
\cite{Vespa} suggest that both belong to the same universality class,
Ben-Hur and Biham \cite{Biham} question these results.

In what follows we implement the sandpile model on the 8-fold quasiperiodic
rhombus-tiling, also called Am\-mann-Been\-ker tiling. Due to the specific 
geometry of qua\-si\-pe\-rio\-dic tilings it can
become either anisotropic with complete non-deterministic toppling rules,
or isotropic with deterministic topplings, or a mixture of both.

\section{Non-periodic tilings and SOC}
\label{sec:1}
\begin{figure}
\resizebox{0.45\textwidth}{!}{%
  \includegraphics{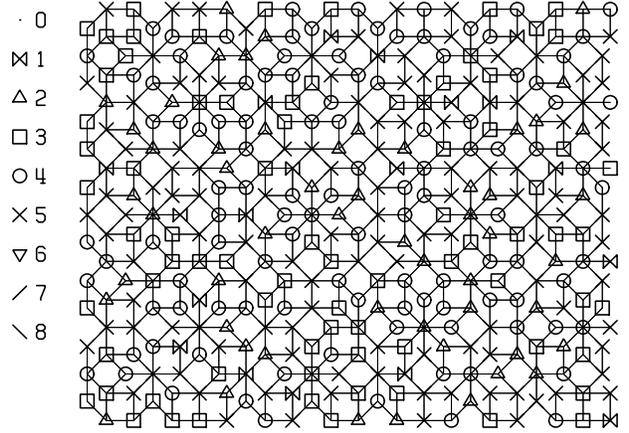}
}
\caption{
The ideal Am\-mann-Been\-ker tiling and a critical state at the end of an
avalanche of type {\bf (I)}. The symbols on top of the vertices code the 
number of grains of sand $z_i$  and $z_c=5$.}
\label{fig1}
\end{figure}

Non-periodic tilings \cite{Gruenbaum} are used to describe the geometry 
of quasicrystals \cite{Shechtman}. They appear in two distinguished classes.
The ideal quasiperiodic tilings are produced by a cut-and-pro\-ject-scheme
from a higher dimensional (minimal) embedding lattice \cite{Kramer,BJKS}.
Their vertices are given by so-called model sets \cite{Moody} and they
show pure Bragg-peak diffraction spectra with non-crystallographic
symmetries (5,7,8,...-fold). Typically, the tilings are built by two or more 
prototiles. Fig.\ \ref{fig1} shows the 8-fold Am\-mann-Been\-ker tiling
consisting of two prototiles, a square and a $\pi/4$ rhombus.

The second class of non-periodic tilings are the so-called
random tilings. They are constructed by the same prototiles
as the ideal ones but this time these prototiles fit together stochastically
in such a way 
that they fill the entire space, face to face without gaps and overlaps
\cite{Elser,Henley}. They also can be embedded into a higher dimensional
lattice but their diffraction properties change. They can keep the same
symmetry of the ideal counterparts but this time, their diffraction
spectrum consists of Bragg-peaks and diffuse scattering (in 3D) or
algebraic peaks (in 2D).

Beneath the long-range correlations which govern the diffraction properties,
the main difference of non-periodic tilings compared to crystallographic
tilings is in their local neighborhood. Whereas we have a fixed translational
invariant neighborhood for crystals, we have a finite atlas of
different vertex configurations without translational invariance for
quasicrystals. For the example of the Am\-mann-Been\-ker tiling, we find
6 different vertices in the ideal and 16 different configurations in the 
random version. The number of nearest neighbors ranges from 3 to 8
\cite{BJ}.

Concerning SOC-models, one now could expect two major influences. 
Both, long-range correlations  and the diversity of the local neighborhoods
of the tilings, could develop a difference in the avalanche distribution.
It is not necessary to test a lot of different quasiperiodic tilings
because the generic features of quasiperiodicity are the same.
Symmetry or the type of the tiling plays a minor r{\^o}le.
Additionally, the local configurations allow different implementations
of e.g.\ the sandpile model. For simplicity we will focus on three
potentially different versions (there are more possible but typically
they fall in the same categories) on the 8-fold Am\-mann-Been\-ker tiling:

\begin{itemize}
\item[(I)]
anisotropic and non-deterministic: The critical height $z_c$ and the number 
of toppling grains $n_i=z_c>0$ are equal for all vertices $i$ of the tiling.
At every toppling step  one has to make a random choice to distribute the
$n_i$ grains among the $N_i$ neighbors $(3 \leq N_i \leq 8)$ of the 
toppling vertex $i$.

\item[(II)] 
partly isotropic and deterministic, partly not: 
$z_c$ is equal for all vertices $i$ but greater or equal to the minimal number
of neighbors $N_{\min}=3$  and less than the maximal number of neighbors 
$N_{\max}=8$. The number of topplings at site $i$ is $n_i=\min(z_c,N_i)$ 
where $N_i$ again is the number of neighbors of vertex $i$. This way, the 
toppling rules are locally isotropic and deterministic for vertices with 
$N_i \leq z_c$ but anisotropic and non-deterministic for all the others because 
of the random choice one has to make in order to distribute the $n_i=z_c$
grains among the $N_i>z_c$ neighbors.

\item[(III)] 
isotropic and deterministic: $z_c$ is greater or equal than the 
maximal number of neighbors $N_{\max}=8$, but the number of toppling grains 
of vertex $i$ is always $n_i=N_i$. This way, every bond will transport the
same amount of grain in both directions and the number of topplings is given
locally by the number of neighbors.

\end{itemize}

After a suitable thermalization, we perform $2*10^6$ ($L=16, 32, 64$)
and $5*10^5$ ($L=128,256$) avalanches for the
different scenarios. Each avalanche is seperated by an appropriate
thermalization time to prevent cross-correlations. The system size 
($L \times L)$ is given by the length $L$ which varies from 16 to 
256 whereas the absolute scale is given by the bond length of the tiling 
that is equal to one. For each setting, the avalanche distributions
are stored. According to the finite size scaling assumption, the data
for the different length scales should collaps to a single curve that
is determined by the critical exponents $\tau_x, \beta_x$. Fig.\ \ref{fig3}
shows the data collaps of the probability distributions of the number of 
topplings $s$ for version {\bf (I)}, Fig.\ \ref{fig4} for version {\bf (II)}, 
and Fig.\ \ref{fig5} for the isotropic deterministic version {\bf (III)}.
Table \ref{tab1} gives the extracted numerical values for the exponents.
The errors given in the brackets are the statictical errors of the fitting
procedure.

\begin{figure}
\resizebox{0.38\textwidth}{!}{%
  \includegraphics{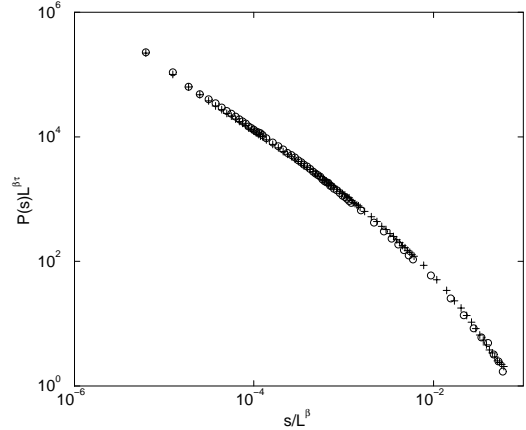}
}
\caption{
Comparsion of anisotropic avalanche distributions of an ideal and
a random tiling version of the Am\-mann-Been\-ker tiling at L=128.}
\label{fig2}
\end{figure}

\begin{figure}
\resizebox{0.4\textwidth}{!}{%
  \includegraphics{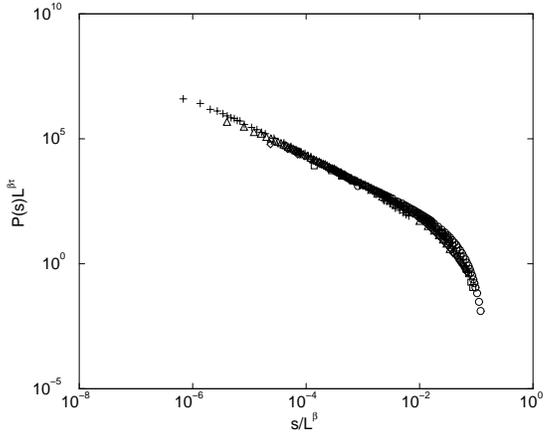}
}
\caption{
Data Collaps of the probability distribution of the number of topplings $s$
for the anisotropic, non-deterministic version {\bf (I)} on the
ideal Am\-mann-Been\-ker tiling.  System lengths are L=16 ($\circ$), 
32 ($\Box$), 64 ($\diamond$), 128 ($\triangle$), and 256 (+).}
\label{fig3}
\end{figure}

\begin{figure}
\resizebox{0.4\textwidth}{!}{%
  \includegraphics{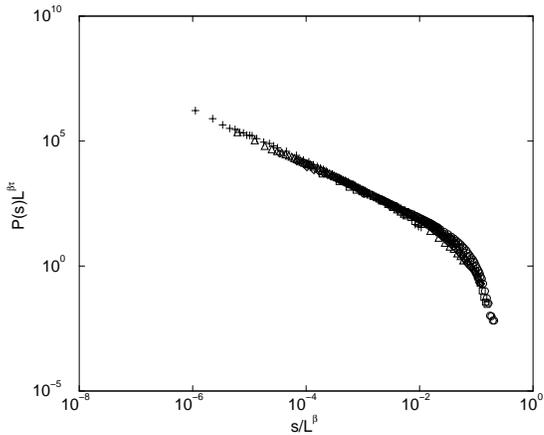}
}
\caption{
Data Collaps of the probability distribution of the number of topplings $s$
for version {\bf (II)} on the ideal Am\-mann-Been\-ker tiling.
System lengths and symbols as in Fig.\ 3.}
\label{fig4}
\end{figure}

\begin{figure}
\resizebox{0.4\textwidth}{!}{%
  \includegraphics{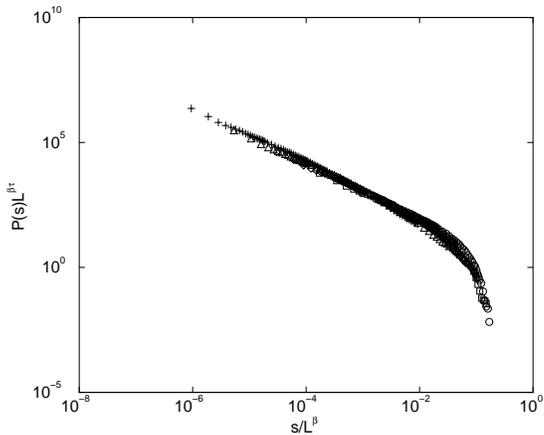}
}
\caption{
Data Collaps of the probability distribution of the number of topplings $s$
for the isotropic, deterministic version {\bf (III)} on the
ideal Am\-mann-Been\-ker tiling.
System lengths and symbols as in Fig.\ 3.}
\label{fig5}
\end{figure}

Within the limits set by the small statistic, all three version
belong to the same universality class: the crtitical exponents and also
the scaling functions seem to be the same. If one compares the
distributions of the ideal quasiperiodic tilings and the corresponding
random tiling, there is no apparent difference (see Fig.\ \ref{fig2}).
Looking at the crystalline counterpart, the 2D square lattice,  the state of 
the art values for the exponents are $\tau_s=1.27(1)$ and $\beta_s=2.73(2)$
\cite{Vespa}. These values are determined by massive parallelized
computation. From the difference to Table \ref{tab1} one cannot conclude
that there exists a different universality class for quasiperiodic graphs.
The values of Table \ref{tab1} are more comparable to older estimates for 
the square lattice based on similar statistics \cite{Grassberger}.
The distributions for the area $a$ and duration $T$ are not shown because
they behave in the same way.

\begin{table}
\caption{The citical exponents $\tau_s, \beta_s$ of the three tested versions
(Statistical errors are given in brackets).}
\label{tab1}
\begin{tabular}{llll}
\hline\noalign{\smallskip}
      & Version {\bf (I)} & Version {\bf (II)}  & Version {\bf (III)}  \\
\noalign{\smallskip}\hline\noalign{\smallskip}
$\tau_s$  & 1.20(3) & 1.19(3) & 1.19(3) \\
$\beta_s$ & 2.5(2)  & 2.5(2)  & 2.5(2) \\
\noalign{\smallskip}\hline
\end{tabular}
\end{table}

\section{Conclusion}

We have performed numerical estimates of the critical exponents of
the 2D sandpile model on the 8-fold quasiperiodic Am\-mann-Been\-ker tiling.
Thereby, three different version (anisotropic-non-deterministic, 
isotropic-deterministic and a mixed one) were implemented on the ideal
and on the random Am\-mann-Been\-ker tiling. The measured exponents are 
comparable 
to early numerical results of the 2D square lattice and might suggest that all
tested version belong to the same universality class as the 2D square lattice.
The problem however remains to substantiate this, e.g.\ by better statistics.
This can only be achieved by a massive use of a parallel computing
which is not at hand at the moment. But the situation seems to be similair
to the investigation of Ising-models \cite{micha} which led to the conlcusion 
that aperiodic and periodic versions belong to the same universality class.

The author wants to thank S.\ Maslov and M.\ Baake for discussion and 
helpful comments.

%
%

\end{document}